\documentclass[final,5p,times,twocolumn]{elsarticle}

\usepackage{epsfig}
\usepackage{amssymb}
\usepackage{amsthm}
\usepackage{amsmath}
\usepackage{hyperref}
\usepackage{bbm}
\usepackage{upgreek}
\usepackage[hyperref]{xcolor}

\hypersetup{
	colorlinks=true,
}
\setcitestyle{square}
\setcitestyle{citesep={,}}

\newcommand{\ket}[1]{\vert{#1}\rangle}

\newcommand{\outpr}[2]{\vert{#1}\rangle\langle{#2}\vert}

\newcommand{\proj}[1]{\outpr{#1}{#1}}

\makeatletter
\def\ps@pprintTitle{%
	\let\@oddhead\@empty
	\let\@evenhead\@empty
	\def\@oddfoot{}%
	\let\@evenfoot\@oddfoot}
\makeatother
\biboptions{numbers,sort&compress}

\begin{document}
\hypersetup{
	linkcolor=red,
	urlcolor=red,
	citecolor=blue
}
\begin{frontmatter}

\title{ Bang-Bang Optimal Control of Large Spin Systems: \\ Enhancement of $^{13}$C-$^{13}$C Singlet-Order at Natural Abundance}
\author[add1]{Deepak Khurana}
\ead{deepak.khurana@students.iiserpune.ac.in}
\author[add1,add2]{T. S. Mahesh}

\address[add1]{Department of Physics and NMR Research center,} 
\address[add2]{	Center for Energy Sciences,
	
	Indian Institute of Science Education and Research, Pune 411008, India }%
\ead{mahesh.ts@iiserpune.ac.in}

\begin{abstract}
Using a Bang-Bang optimal control (BB) technique, we transfer polarization from abundant high-$\gamma$ nuclei directly to singlet order.  This approach is analogous to  algorithmic cooling (AC) procedure used in quantum state purification.  Specifically, we apply this method for enhancing the singlet order in a natural abundant $^{13}$C- $^{13}$C spin pair using a set of nine equivalent protons of an 11-spin system. Compared to the standard method not involving polarization transfer, we find an enhancement of singlet order by about three times.  In addition, since the singlet magnetization is contributed by the faster relaxing protons, 
the recycle delay is halved.
Thus effectively we observe a sensitivity enhancement by 4.2 times or a reduction in the overall experimental time by a factor of 18.  We also discuss a possible extension of AC, known as heat-bath algorithmic cooling (HBAC).  

\end{abstract}


\end{frontmatter}
\section{Introduction}
Not many experimental architectures allow as elaborate control on quantum dynamics as that of NMR.  Several powerful RF control techniques such as composite pulses \cite{levitt1986composite}, adiabatic pulses \cite{tannus1997adiabatic}, band-selective/broadband pulses \cite{geen1991band,wimperis1994broadband,skinner2003application,feike1996broadband} are being routinely used in NMR spectroscopy.  Numerical methods such as 
strongly modulating pulses \cite{fortunato2002design},
GRadient Ascent Pulse Engineering (GRAPE) \cite{khaneja2005optimal}, Krotov \cite{vinding2012fast}, etc have also been used for specific purposes in spectroscopy as well as quantum information.  Here we describe an application of Bang-Bang (BB) optimal control that utilizes a sequence of full-power RF pulses with variable phases separated by variable delays \cite{morton2006bang,bhole2016steering,zhou2011bang}.  Generally, the numerical complexity of optimal control techniques scales rapidly with the size of the spin system, thus limiting their applications.  On the other hand, BB relies on one-time matrix exponentiation to build basic unitaries and hence it's complexity scales much slower, and therefore is applicable also for fairly larger spin systems \cite{bhole2016steering}.  In this work we utilize the BB control to directly transfer polarization from a set of ancillary spins to the long-lived singlet-order in a spin-pair.  

Right after its conception, long-lived singlet-order has gained significant theoretical and experimental interest \cite{carravetta2004beyond,pileio2010storage,carravetta2004long,carravetta2005theory,pileio2009theory,levitt2012singlet,pileio2012long,pileio2008long,emondts2014long,feng2012accessing,ahuja2009long,devience2013preparation,pileio2007j,tayler2011singlet,dumez2014long,grant2008long,feng2014long,claytor2014accessing,kadlecek2010optimal} due to its wide range of applications such as study of slow molecular processes \cite{sarkar2007singlet}, characterizing molecular diffusion \cite{sarkar2008measurement,cavadini2005slow}, precision measurement of scalar couplings \cite{pileio2009extremely}, obtaining molecular structure information \cite{tayler2010determination}, and storage of hyper polarization \cite{vasos2009long,warren2009increasing,pileio2013recycling,ahuja2010proton}. 

In a pair of two-level quantum particles with individual basis states $\{\ket{\uparrow},\ket{\downarrow}\}$,  antisymmetric singlet state is
\begin{equation}
\ket{S_0} = \frac{\ket{\uparrow\downarrow}-\ket{\downarrow\uparrow}}{\sqrt{2}}
\end{equation}
and symmetric triplet states are
\begin{align}
\ket{T_0} &= \frac{\ket{\uparrow\downarrow}+\ket{\downarrow\uparrow}}{\sqrt{2}} \nonumber \\
\ket{T_+} &= \ket{\uparrow\uparrow} ~\mbox{and}, \nonumber \\
\ket{T_-} &= \ket{\downarrow\downarrow}.
\end{align}

In NMR, the basis states are usually the spin eigenstates in the Zeeman magnetic field, and under normal conditions it is hardly possible to prepare any pure quantum state.  However, it is often possible to prepare an excess population in one of the quantum states relative to uniformly populated remaining  states. Thus an excess population of singlet state relative to a uniformly populated triplet states is represented by the density operator
\begin{equation}
\rho_S = (1-\epsilon_S) \mathbbm{1}_4/4 + \epsilon_S \proj{S_0},
\end{equation}
where $\mathbbm{1}_4$ is the four-dimensional identity operator and the scalar quantity $\epsilon_S$ quantifies  the singlet-order \cite{levitt2012singlet}.
Since the dominant intra-pair  dipolar relaxation process does not connect subspaces of different symmetries, the singlet-order often  lives much longer than other non-equilibrium states whose lifetimes are limited by the spin-lattice relaxation time constant $T_1$  \cite{carravetta2005theory,pileio2009theory}.  In favorable cases, singlet life-times as long as over 50 times $T_1$ have also been observed \cite{dumez2014long}.

One way to access singlet-order is to  utilize the chemical shift separation (along with J-coupling) between two spins to prepare a mixture $\proj{S_0}-\proj{T_0}$ of singlet and  triplet states.  This is followed by suppression of chemical shift to impose symmetry, achieved either with    low-field switching by shuttling the sample out of the magnet \cite{carravetta2004beyond}, or with a strong RF spin-lock while retaining the high-field \cite{carravetta2004long}.  
After a desired storage period, the chemical shift separation is restored and the singlet-order is converted back into observable single quantum coherence.

Later, accessing singlet-order in systems with chemical equivalence, but magnetic inequivalence w.r.t. a chemically equivalent ancillary spin-pair, was discovered \cite{feng2012accessing}.  In this case, each of the chemically equivalent spin-pairs exist in singlet states at high magnetic fields without requiring external spin-lock to impose symmetry.  It was also shown that by exploiting the higher sensitivity of ancillary $^{1}$H-$^{1}$H spin-pair, one can prepare, store, and detect $^{13}$C-$^{13}$C singlet order either with isotopic labeling \cite{feng2013storage} or even at natural abundance \cite{claytor2014measuring}.  

In this work, we show that using the BB optimal control techniques, we can directly transfer polarization from ancillary protons to enhance naturally abundant $^{13}$C-$^{13}$C singlet order.  This method is widely applicable in a variety of systems where a pair of spins with or without chemical equivalence are coupled to a few ancillary spins.

Although the concept of polarization transfer has long been a part of NMR spectroscopy \cite{morris1979enhancement,pines1972proton}, it has been revisited in quantum information while attempting to achieve a small set of highly pure quantum bits (system qubits) at the expense of purity of a large number of ancillary quantum bits (reset qubits).  This process known as algorithmic cooling (AC) systematically transfers entropy from system qubits to reset qubits \cite{schulman1999molecular,park2016heat}. 
Motivated by these concepts, we refer to the single iteration polarization transfer as AC.
Heat bath algorithmic cooling (HBAC) is a nonunitary extension of AC, that involves removal of the extra entropy from the reset qubits to an external bath so that AC can be iteratively applied to achieve higher purity of system qubits \cite{boykin2002algorithmic}.

The paper is organized as follows: In section \ref{method}  we describe BB optimal control in detail, along with our spin system and pulse sequence. In section \ref{exp} we describe experimental results and simulations. Finally, section \ref{con} contains   discussions and conclusions.

\section{Methods}
\label{method}
\subsection{Bang Bang (BB) optimal control}
  Consider a system in a state $\rho_\mathrm{in}$ that needs to be steered to a target state $\rho$.  We discretize the time evolution into $N$ segments each of duration $\Delta t$. In the  rotating-frames of the RF carriers,
 let ${\cal H}_0$

 be the internal Hamiltonian of the system and
  \begin{eqnarray}
  {\cal H}_{k,n} &=& A_{k,n} \left(I_{x}^k \cos\phi_{k,n} + I_{y}^k\sin\phi_{k,n}\right) \nonumber \\
  &=& A_{k,n} Z_{k,n} I_{x}^k Z_{k,n}^\dagger
  \end{eqnarray}
  be the RF Hamiltonian on $k^\mathrm{th}$ channel and $n^\mathrm{th}$ segment.  Here $A_{k,n},~\phi_{k,n}$ are the amplitudes and phases respectively,   $Z_{k,n} = \exp(-i \phi_{k,n} I_{z}^k)$, and
  $I_{x}^k$, $I_{y}^k$, $I_{z}^k$ are the spin operators on $k^\mathrm{th}$ nuclear species.
  The full piecewise continuous Hamiltonian 
  \begin{equation}
  {\cal H}_n = {\cal H}_0 + \sum_k {\cal H}_{k,n}
  \end{equation}
  achieves an effective unitary evolution $U = \prod_{n=1}^{N}e^{-i{\cal H}_n \Delta t}$.

        \begin{figure}
        	\includegraphics[trim = 0cm 5cm 1cm 1cm, clip, width=8.5cm]{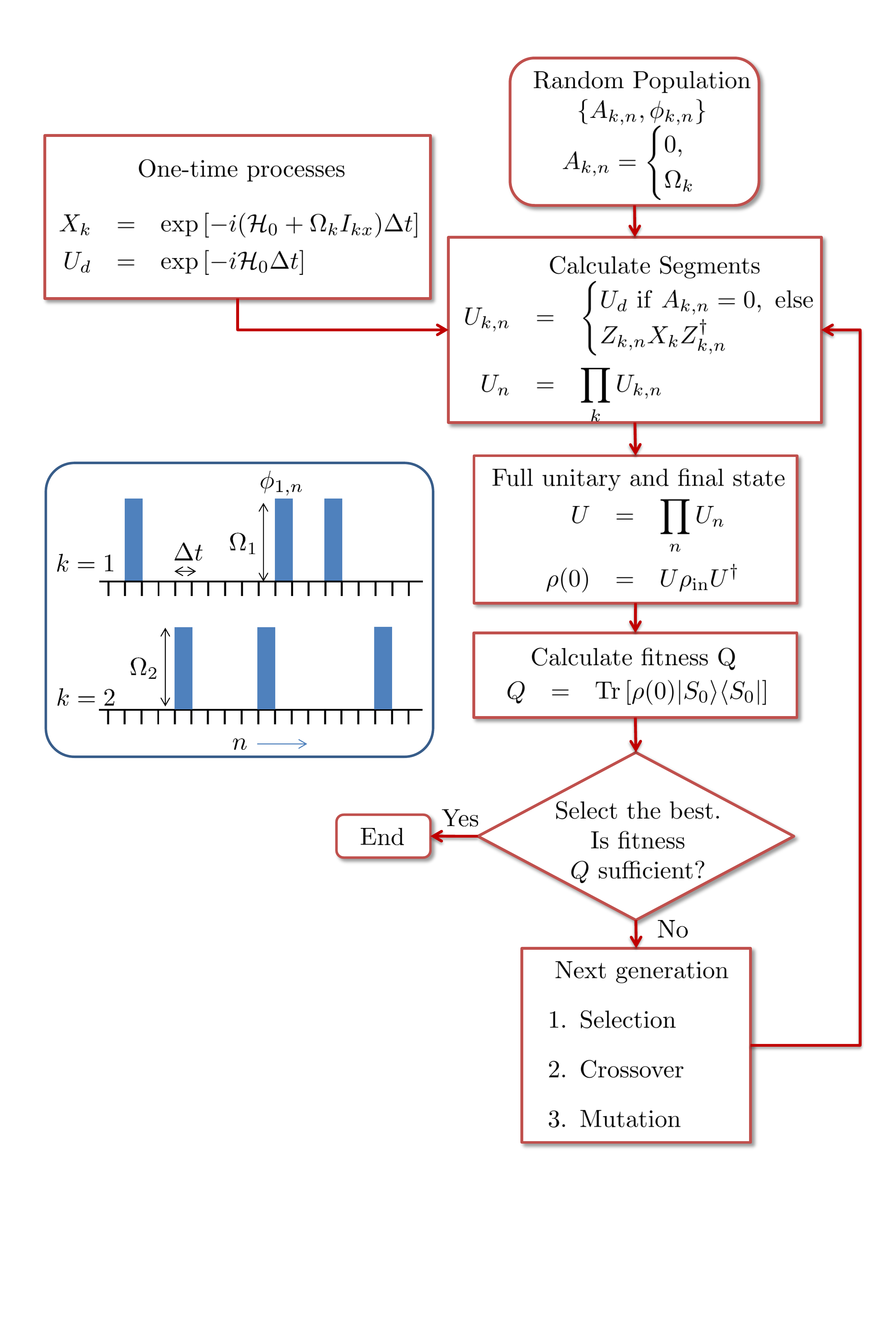}
        	\caption{The flowchart describing BB optimal control with genetic algorithm.  A schematic diagram of BB control is shown in the inset.  Here each rectangular box corresponds to a full power RF pulse, called a \textit{bang}.}
        	\label {bb}
        \end{figure}

  In this work, we try to find a unitary $U$ that  prepares the target state 
  \begin{eqnarray}
  \rho(0) &=& U \rho_\mathrm{in} U^\dagger \nonumber \\
  &=&  \left\{1-\epsilon_{S}(0)-\epsilon_\Delta\right\}\frac{\mathbbm{1}_4}{4} + \epsilon_{S}(0) \proj{S_0}+ \epsilon_\Delta \rho_\Delta,
  \end{eqnarray}
 containing a 
   long-lived singlet component $\proj{S_0}$ with a maximum singlet-order $\epsilon_{S}(0)$. Here $\rho_\Delta$ is an undesired, though unavoidable,  component  containing  triplet states as well as other artifact coherences.   The expectation value of the singlet component $\proj{S_0}$ in $\rho(0)$ is
 \begin{equation}
 Q = \mathrm{Tr}\left[\rho(0)~ \proj{S_0}\right] = \frac{3\epsilon_S(0)+1}{4}.
 \end{equation}
Therefore by maximizing $Q$ via BB control, we can obtain an unitary preparing a maximum singlet-order.
A  subsequent spin-lock of duration $\tau$  rapidly damps out the short-lived component  $\rho_\Delta$ towards the maximally mixed state ${\mathbbm{1}_4}/{4}$, such that the purified state is of the form
\begin{equation}
\rho(\tau) = \left\{1-\epsilon_S(\tau) \right\} \mathbbm{1}_4/4 + \epsilon_S(\tau) \proj{S_0},
\end{equation}  
where 
$ \epsilon_{S}(\tau) =   \epsilon_{S}(0) e^{-\tau/T_{S}}$ is the singlet-order decayed due to the long singlet life-time $T_{S}$. In the next section we use AC to enhance the singlet-order from $\epsilon_S$ to $\epsilon_S^\mathrm{AC}$ by polarization transfer from ancillary spins.

   As opposed to schemes like GRAPE \cite{khaneja2005optimal} which use smooth RF modulations,
   BB control employs pulses having either zero or full RF amplitudes ($A_{k,n} = \Omega_{k}$ or $0$) but variable phases ($\phi_{k,n}$) to generate arbitrary unitaries.  A flowchart describing various steps of the BB optimal control using genetic algorithm is shown in Fig.\ref{bb}.  The major advantage of BB control is that the exponentiation of Hamiltonian to obtain the basic unitary ($X_k$) as well as the delay unitary ($U_d$)  is rendered an one-time process that is outside of the iterations.  Matrix exponentiation is a bottleneck in conventional algorithms based on amplitude modulation, particularly for large spin systems.  The unitaries corresponding to arbitrary bangs are obtained by rotating the basic opertor $X_k$ about $\hat{z}$ axis, i.e., $U_{k,n} = Z_{k,n} X_k Z_{k,n}^\dagger$.  Here $Z_{k,n}$ is a diagonal operator in Zeeman basis and hence is efficiently computed during the run-time of iterations.  Thus BB method allows quantum control of large spin systems as demonstrated in the later section.  It is even more efficient in designing RF sequences with low duty-cycle requiring long evolutions of internal Hamiltonian such as polarization transfer operations.

\subsection{Spin System}
\begin{figure}[h]
	\includegraphics[trim = 3cm 7cm 3cm 3cm, clip, width=8cm]{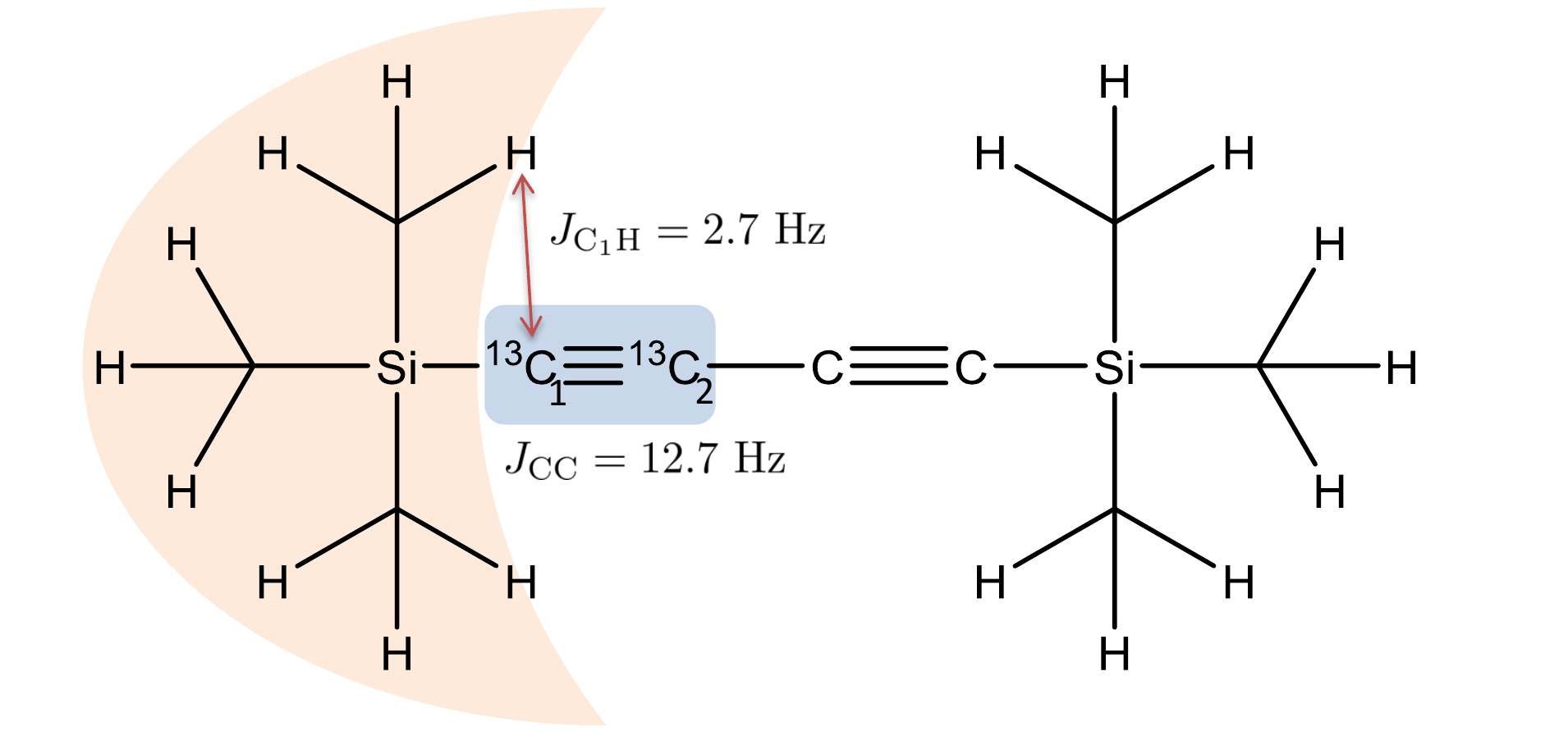}
	\caption{Structure of  $1,4$-Bis(trimethylsilyl)butadiyne. Here protons in the shaded area act as ancillary spins which provide polarization to $^{13}$C-$^{13}$C singlet-order.}
	\label {molecule}
\end{figure}

To demonstrate HBAC, we use an 11-spin system including a pair of naturally abundant, weakly-coupled $^{13}$C spins surrounded by nine chemically equivalent $^1$H spins of $1,4$-Bis(trimethylsilyl)butadiyne (BTMSB).  The sample was prepared by dissolving 120 mg of BTMSB in 0.7 ml of CDCl$_{3}$  (0.88 M).
We use the protons to directly prepare enhanced $^{13}$C-$^{13}$C singlet-order. 
  The molecular structure of BTMSB is shown in Fig. \ref{molecule}.  The molecular symmetry provides 
twice the probability of naturally abundant $^{13}$C-$^{13}$C pairs. 
The chemical shift difference between the two $^{13}$C spins is $2.32$ ppm, and the $^{13}$C$_1$-$^{13}$C$_2$ J-coupling constant is 12.7 Hz, while J-coupling between $^{13}$C$_1$ and the closest equivalent protons is 2.7 Hz. The spin-lattice relaxation time constants ($T_1$) are about 3 s, 6.5 s, and 8.2 s for $^1$H, $^{13}$C$_1$ and $^{13}$C$_2$ respectively.  The effective transverse relaxation time constants ($T_2^*$) are respectively 0.3 s, 2.5 s, and 2.9 s.

 \subsection{Pulse sequence}
\begin{figure}[b]
	\includegraphics[trim =2cm 6cm 2cm 1cm, clip, width=9cm]{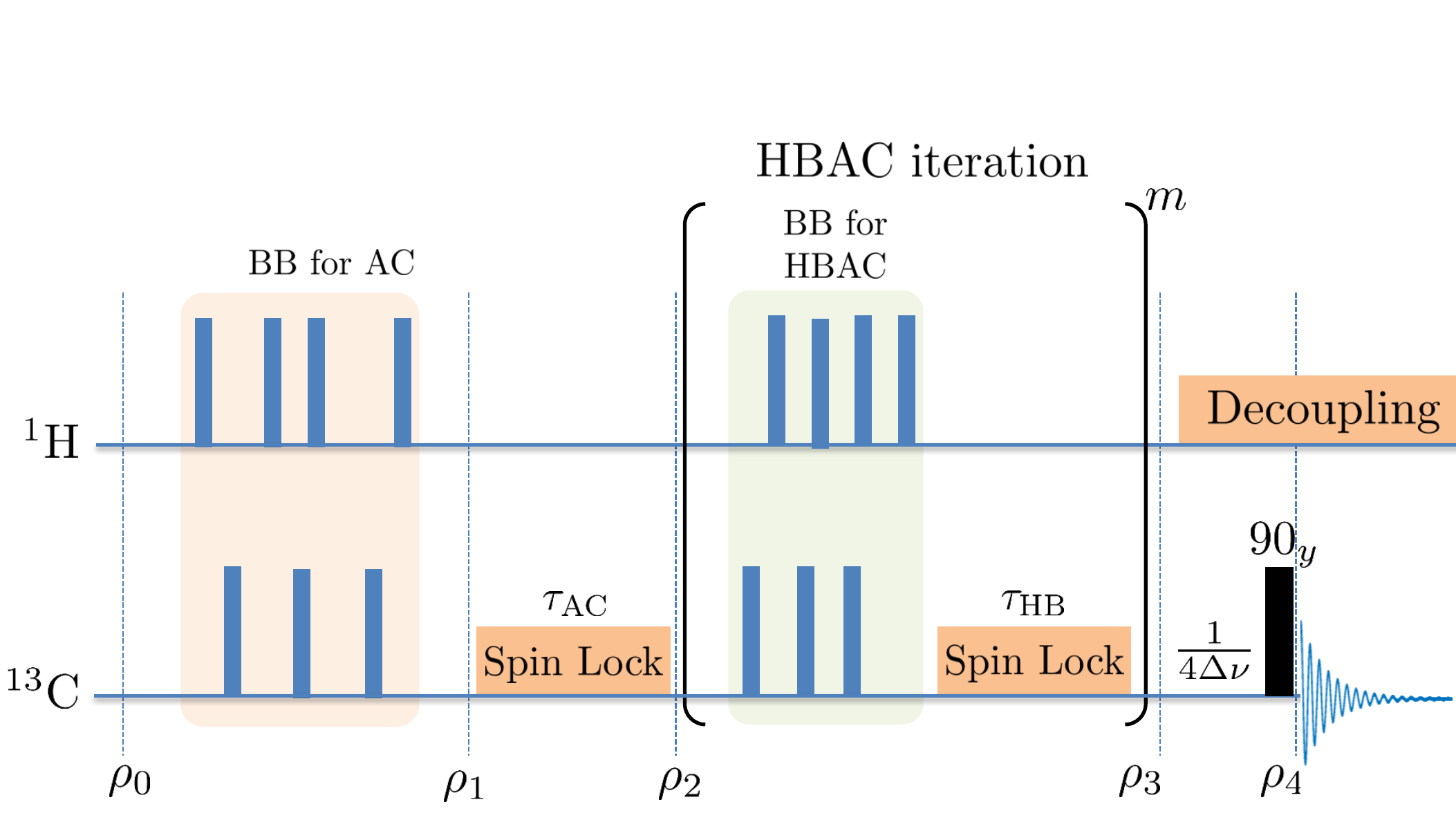}
	\caption{Pulse sequence for the preparation and enhancement of $^{13}$C-$^{13}$C singlet-order using $^1$H spins. Here $m$ can take integral values $0,1,2,\cdots$ etc.}
	\label {Acpulse}
\end{figure}

The pulse sequence employed for the preparation and enhancement of $^{13}$C-$^{13}$C singlet polarization is shown in Fig \ref{Acpulse}.  The initial thermal equilibrium state of the system is
\begin{equation}
 \rho_{0} = \frac{\mathbbm{1}_2^{\otimes 11}}{2^{11}}+ \epsilon_\mathrm{C} \left( I_z^{\mathrm{C}_1} +
I_z^{\mathrm{C}_2}\right)
+ \epsilon_\mathrm{H} \sum_{j=1}^{9} I_z^{\mathrm{H}_j}
\label{rho0}
 	\end{equation}
 where $\epsilon_\mathrm{C}$ and $\epsilon_\mathrm{H}$ are the carbon and proton polarizations respectively, and
 $\epsilon_\mathrm{H}/\epsilon_\mathrm{C} = \gamma_\mathrm{H}/\gamma_\mathrm{C} \simeq 4$.

Then a BB sequence is applied to prepare $^{13}$C-$^{13}$C singlet-order.  Thus the reduced density operator for carbon spins is now
 \begin{equation}
 \rho_{1} = \left[1-\epsilon_{S}^\mathrm{AC}(0)-\epsilon_\Delta\right]\frac{\mathbbm{1}_4}{4} + \epsilon_{S}^\mathrm{AC}(0) \proj{S_0}+ \epsilon_\Delta \rho_\Delta,
 \label{zerospinlock}
 \end{equation}
where $\epsilon_{S}^\mathrm{AC}(0)$ represents the enhanced singlet-order. At the end of the spin-lock of duration $\tau_\mathrm{AC}$, one obtains a high quality singlet state
\begin{equation}
\rho_{2} = \left[1-\epsilon_{S}^\mathrm{AC}(\tau_\mathrm{AC})\right]\frac{\mathbbm{1}_4}{4} + \epsilon_{S}^\mathrm{AC}(\tau_\mathrm{AC}) \proj{S_0},
\end{equation}
with the singlet-order $\epsilon_{S}^\mathrm{AC}(\tau_\mathrm{AC})$.

Suppose $^1$H spins have much shorter $T_1$ relaxation time constant compared to the life time of singlet ($T_{S}$).  Then during the spin-lock duration, $^1$H spins regain polarization by spin-lattice relaxation and are available for further polarization transfer to $^{13}$C-$^{13}$C singlet state.
In our pulse sequence this is achieved by another  BB pulse. This process (Fig. \ref{Acpulse}) known as HBAC can be iterated to further enhance the singlet-order in favorable systems.  At the end of  $m$ HBAC iterations we obtain the state
\begin{equation}
\rho_{3} = \left[1-\epsilon_{S}^{\mathrm{HB}_m}(\tau_\mathrm{HB})\right]\frac{\mathbbm{1}_4}{4} + \epsilon_{S}^{\mathrm{HB}_m}(\tau_\mathrm{HB}) \proj{S_0},
\label{rho4}
\end{equation}
where $\epsilon_{S}^{\mathrm{HB}_m}(\tau_\mathrm{HB})$ is the singlet-order after the final spin-lock of duration $\tau_\mathrm{HB}$.

Finally, we convert the singlet polarization into,
\begin{equation}
\rho_{4} = \frac{\mathbbm{1}_4}{4} - \epsilon_{S}^{\mathrm{HB}_m}(\tau_\mathrm{HB}) \left(
I_z^{\mathrm{C}_1}I_y^{\mathrm{C}_2}-
I_y^{\mathrm{C}_1}I_z^{\mathrm{C}_2}+
I_x^{\mathrm{C}_1}I_x^{\mathrm{C}_2}
\right),
\end{equation}
where $zy$ and $yz$ terms form the observable single quantum coherences of $^{13}$C spins.
In the following section we describe experimental results and numerical analysis.

\section{Experimental results and numerical analysis}
\label{exp}
All experiments were performed using a 9.4 T (400 MHz) Bruker NMR spectrometer at an ambient temperature of 298 K using a standard high-resolution BBO probe. The transition-selective Gaussian pulse was of 750 ms duration. The BB pulse for AC was of duration 296 ms and that of HBAC was of 248.5 ms.

\begin{figure}[h]
	\begin{center}
		\includegraphics[trim = 1cm 0.5cm 1cm 0cm, clip, width=8.5cm]{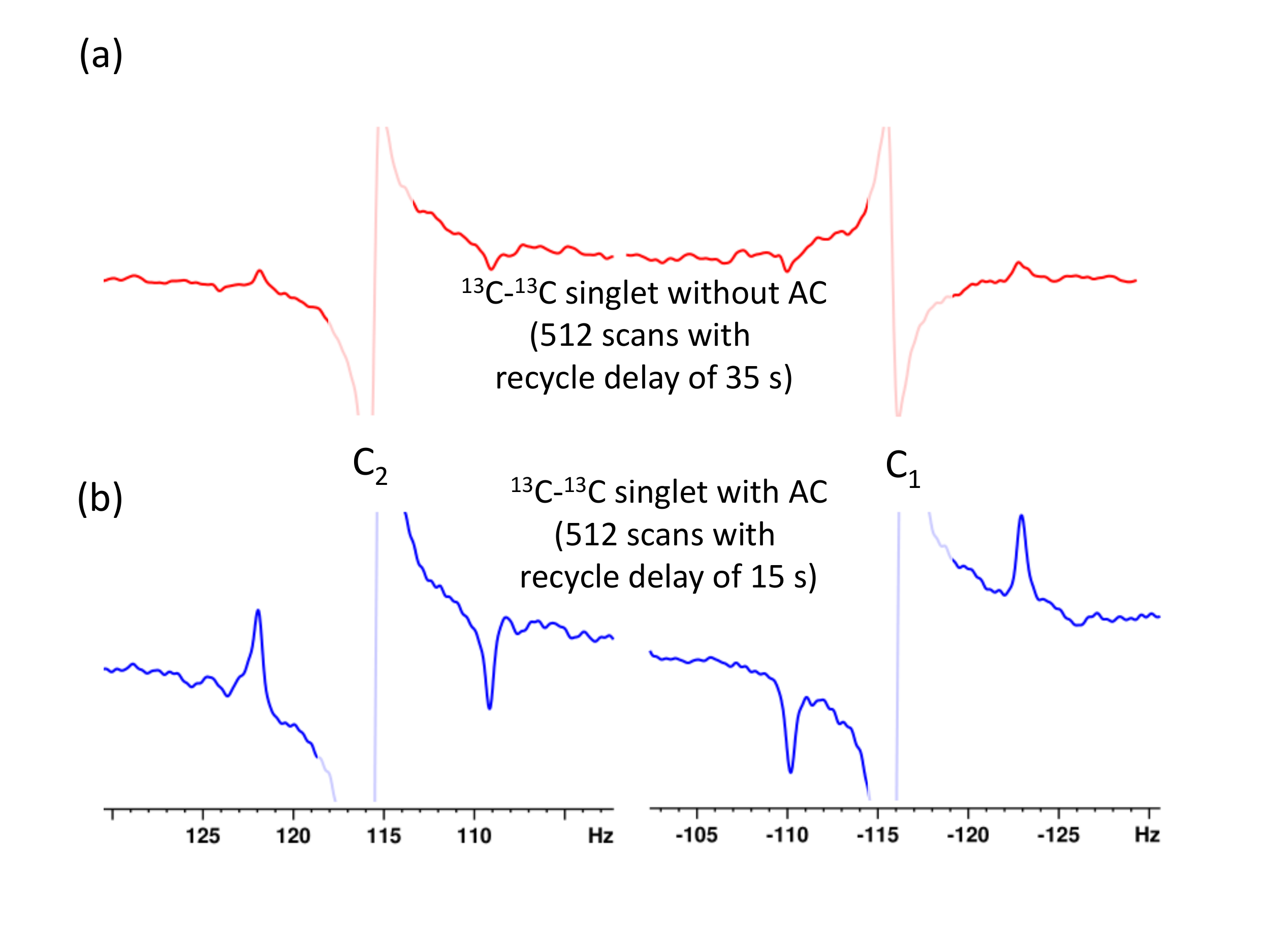}
		\caption{$^{13}$C spectra of BTMSB obtained by converting  $^{13}$C-$^{13}$C singlet-order at natural abundance into single-quantum coherence:  (a) without AC and  (b) with AC. Both  spectra were recorded with WALTZ-16 spin-lock (1.5 kHz, $\tau_\mathrm{AC} = 10$ s).  The central peaks corresponding to $^{13}$C-$^{12}$C pairs are de-emphasized.}
		\label{ACnoAC}
	\end{center}
\end{figure}

Fig. \ref{ACnoAC} (a) displays the $^{13}$C spectra corresponding to $^{13}$C-$^{13}$C singlet-order at natural abundance without AC, and was obtained with the standard sequence without involving any polarization transfer  \cite{carravetta2004long}.   The recycle delay was set to 35 s (approximately five times $T_1$ of carbons) and a total 512 scans were recorded.  Although the characteristic signature of the singlet state is visible in terms of the antiphase magnetizations, the signal to noise ratio is rather poor.

Fig. \ref{ACnoAC} (b) displays the $^{13}$C spectra corresponding to $^{13}$C-$^{13}$C singlet-order  with AC again recorded with 512 scans.  Since the polarization is mainly contributed by $^1$H spins, we need a recycle delay of only 15 s (approximately five times $T_1$ of protons) and accordingly required only half the experimental time as that of without AC.  Both spectra in Fig. \ref{ACnoAC} were recorded with the same spin-lock duration $\tau_\mathrm{AC}$ of 10 s.
However, the signal to noise ratio with AC is about twice that of without AC spectrum. The estimated enhancement of singlet-order with AC compared to that of without AC, i.e., $\epsilon_S^{\mathrm{AC}}/\epsilon_S$, is about 3.

\begin{figure}
	\begin{center}
	\includegraphics[trim = 1cm 0.5cm 2cm 1cm, clip, width=8cm]{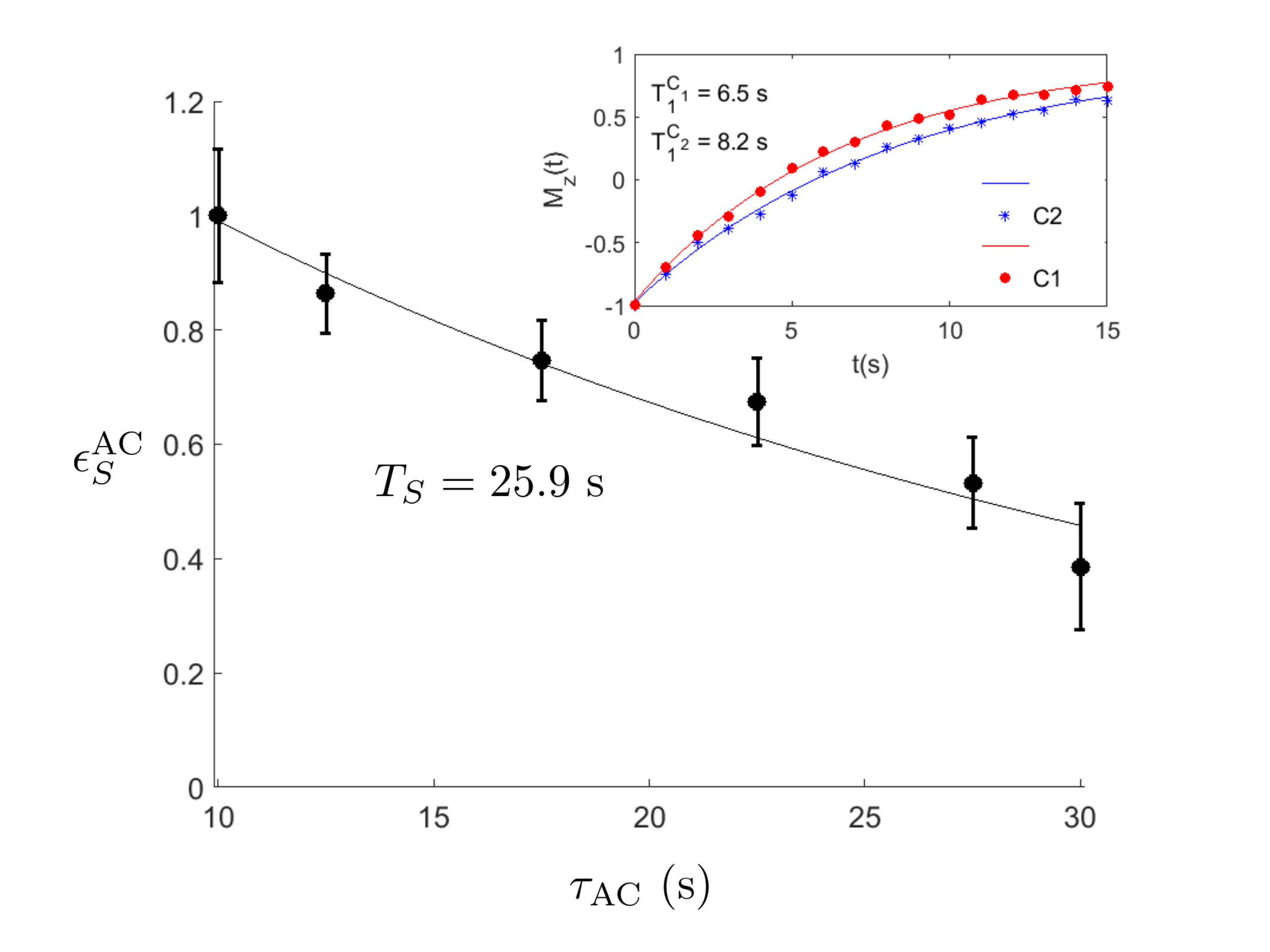}
	\caption{The decay of singlet polarization versus spin-lock duration $\tau_\mathrm{AC}$. The vertical axis is normalized w.r.t. the first data point. Inversion-recovery  curves and corresponding $T_1$ values of both carbons are shown in the inset. }
	\label{singletdecay}
	\end{center}
\end{figure}

The enhanced singlet-order allows us to conveniently monitor its decay versus the spin-lock duration $\tau_\mathrm{AC}$.  The results shown in 
Fig. \ref{singletdecay} indicate the singlet decay constant $T_S$ of about 25.9 s.  Thus,
the singlet-order is approximately $3$ to $4$ times longer lived compared to the $T_1$ values of carbons.

In this particular spin system, we did not observe any advantage of HBAC over AC.  
HBAC is suitable for systems with fast relaxing  ancillary spins and very slow relaxing system spins \cite{pande2017strong}.  In such a system, protons recover their magnetization (after AC) much faster than the decay of singlet state, so that further polarization transfer can be carried out.  In our system, the $T_1$ to $T_S$ contrast was insufficient to observe this effect.

\begin{figure}[b]
	\includegraphics[trim = 0cm 1.8cm 0cm 1.8cm, clip, width=9.5cm]{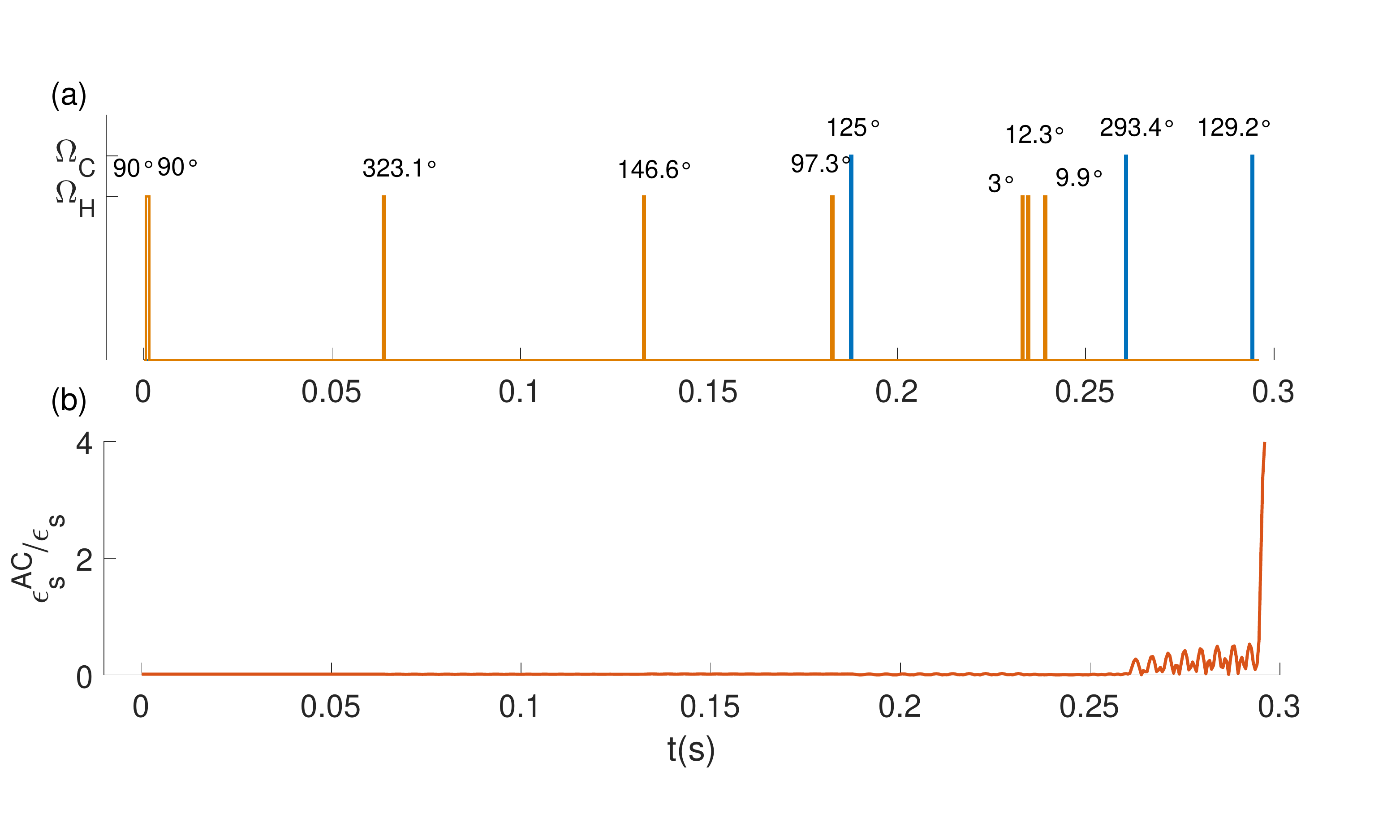}
	\caption{(a) BB profile of obtained for AC.   It consists of bangs
		on proton as well as carbon channels with amplitudes $\Omega_\mathrm{H}$ and $\Omega_\mathrm{C}$ respectively.  Each bang is of duration 0.5 ms.  The phases in degrees are shown above the bangs. (b) Progress of enhancement factor $\epsilon^\mathrm{AC}_S / \epsilon_S$ versus time during the BB sequence.	}
	\label{Fpac}
\end{figure}

We now numerically analyze the BB sequences to understand the dynamics of singlet-order enhancement.  Fig. \ref{Fpac} (a) shows the BB profile of AC pulse and \ref{Fpac} (b) displays the evolution of singlet-order as a function of time starting from state $\rho_0$ of Eqn. \ref{rho0}.   Here time discretization was done with $\Delta t = 500~\upmu$s, and RF amplitude $\Omega_\mathrm{C}/(2\pi) = 250$ Hz.  Thus each bang corresponds to a $45^\circ$ nutation.

  At the end of the AC sequence the singlet enhancement factor reaches a maximum value of 4.  Experimentally, however, we achieved an enhancement of about 3, presumably due to  RF inhomogeneity, hardware non-linearity, and relaxation effects.

\section{Discussions and conclusions}
\label{con}
Sophisticated quantum control techniques are recently being used in both spectroscopy as well as in quantum information to achieve complex and precise spin dynamics \cite{khaneja2001time,khaneja2005optimal,bhole2016steering,yuan2015time,tovsner2009optimal,fortunato2002design,morton2006bang}.  
The challenge in many of such techniques is the numerical complexity involved in evaluating and optimizing propagators of large spin systems.  In this regard, the Bang-Bang (BB) quantum control technique offers a unique advantage, since it only needs one-time evaluation of basic propagators by matrix exponentiation. Therefore, we can synthesize BB controls for larger spin-systems. Here we have described the various steps in the BB control technique using a flowchart.

In this work, we achieve the quantum control of 11-spin system by transferring polarization from nine ancillary spins into the singlet-order of a spin-pair.  We experimentally demonstrate this method in a naturally rare $^{13}$C-$^{13}$C spin-pair, with a probability of 0.011\%, and obtain an enhanced singlet-order by a factor of 3, compared a standard method without involving polarization transfer.   However, owing to the faster $T_1$ relaxation of the ancillary protons, the BB approach needed only half the experimental time compared to the latter.  Thus effectively, we gain sensitivity enhancement by about 4.2 times or effectively over 18 times reduction in experimental time.

Exploiting the enhanced sensitivity, we investigated the decay of the singlet-order under spin-lock and found it to be three to four times longer lived compared to individual spin-lattice relaxation time constants.

We also investigated the heat-bath algorithmic cooling (HBAC) which attempts to further enhance the singlet-order by iterative transfer of polarization from ancillary spins.  HBAC is particularly suited for systems with fast relaxing ancillary spins and slow relaxing target spins \cite{pande2017strong}.  In principle, the long-lived singlet states are ideal for storing the spin-order between the iterations where ancillary spins re-thermalize by giving away extra heat to their bath.  With this motivation, we explored HBAC in the 11-spin system described above.  However, due to insufficient contrast between the life-times of singlet-order and ancillary spins, as well as insufficient enhancement by each iteration, we could not observe any significant advantage of HBAC process in this system.

The methods described here are can be applied to other homonuclear spin-pairs such as naturally rare $^{15}$N-$^{15}$N or even naturally abundant  $^{31}$P-$^{31}$P pairs.
We also anticipate to find many other interesting applications of BB control techniques in spectroscopy as well as in quantum information processing.
 
\section{Acknowledgments}
We thank Varad Pande for  discussions in algorithmic cooling and Gaurav Bhole for  discussions in BB optimal control. This work was supported by DST/SJF/PSA-03/2012-13 and CSIR 03(1345)/16/EMR-II.

\section*{References}
  \bibliographystyle{elsarticle-num} 
  \bibliography{sp}
\end{document}